\documentclass[amstex,twocolumn,showpacs,floats,floatfix,superscriptaddress,aps,pra]{revtex4}

\usepackage{amssymb}
\usepackage{amsmath}
\usepackage{calc}
\usepackage{graphicx}
\usepackage{bm}
\usepackage{animate}

\usepackage[colorlinks]{hyperref}
\hypersetup{
   colorlinks = true,
   linkcolor=blue,
   citecolor=blue}

\def\be{ \begin{equation} }
\def\ee{ \end{equation} }
\def\bea{ \begin{eqnarray} }
\def\eea{ \end{eqnarray} }
\def\bse{ \begin{subequations} }
\def\ese{ \end{subequations} }
\def\Nb{N_b}
\def\Nn{N_n}
\def\B{\mathcal{B}}
\def\N{\mathcal{N}}
\def\phaseN{\nu}
\def\phaseB{\beta}

\begin{document}

\author{Elica Kyoseva}
\affiliation{Engineering Product Development, Singapore University of Technology and Design, 20 Dover Drive, 138643 Singapore}
\affiliation{Department of Nuclear Science and Engineering, Massachusetts Institute of Technology, 77 Massachusetts Avenue, Cambridge, Massachusetts 02139, USA}
\author{Nikolay V. Vitanov}
\affiliation{Department of Physics, St. Kliment Ohridski University of Sofia, 5 James Bourchier blvd, 1164 Sofia, Bulgaria}

\title{Arbitrarily accurate passband composite pulses for dynamical suppression of amplitude noise}
\date{\today }

\begin{abstract}
We introduce high-fidelity passband (PB) composite pulse sequences constructed by concatenation of recently derived arbitrarily large and arbitrarily accurate  broadband $\B$ and narrowband $\N$ composite sequences.
Our PB sequences allow us to produce flexible and tunable nearly rectangular two-state inversion profiles as a function of the individual pulse area because the width and the rectangularity of these profiles can be adjusted at will.
Moreover, these PB sequences suppress excitation around pulse area 0 and $2\pi$, and suppress deviations from complete population inversion around pulse area $\pi$ to arbitrarily high orders.
These features makes them a valuable tool for high-fidelity qubit operations in the presence of relatively strong amplitude noise.
We construct two types of PB pulses: $\N(\B)$, in which a broadband pulse is nested into a narrowband pulse, and $\B(\N)$, in which a narrowband pulse is nested into a broadband pulse; the latter sequences deliver narrower profiles.
We derive exact analytic formulas for the composite phases of the PB pulses and exact analytic formulas for the inversion profiles.
These formulas allow an easy estimation of the experimental resources needed for any desired qubit inversion profile.
\end{abstract}
\maketitle


\section{Introduction}

Composite pulse sequences were originally developed for use in nuclear magnetic resonance (NMR) to tackle a range of errors which arise due to imperfections in the radiofrequency control fields used to manipulate spin systems \cite{Levitt79,Freeman80,Levitt86,Freeman,Tycko85,Cho86,Wimperis94,Cummins}. Due to their robustness and precision in coherent control of quantum systems composite pulse sequences have further found applications in atomic physics \cite{Torosov,TorosovPRL,Vitanov}, magnetometry with solid state quantum sensors \cite{Aiello}, molecular spectroscopy \cite{ms}, and atom interferometry \cite{josa}. More recently, the advancement of quantum information processing (QIP) posed the requirement of performing extremely accurate controlled unitary operations on quantum systems \cite{Nielsen}. In the experimental realization of QIP errors arise from control field amplitude noise, spatial field inhomogeneities, uncontrolled frequency offset, unwanted chirp, etc., which are known as systematic errors to distinguish them from random noise errors due to decoherence. Composite pulse sequences found further applications as a robust and versatile tool to tackle these and other systematic errors \cite{2,Ivanov2, Jones13}.

Composite pulse technique replaces a single pulse conventionally used to drive a two-state quantum transition by a finite train of pulses with well-defined relative phases. The phases are used as a control tool and can be determined such as to cancel out a systematic error inherent in the single pulse rotations. Thus the propagator of the system can be shaped in a desired manner to create a robust unitary operation, an objective which is impossible with a single pulse or adiabatic techniques. Such composite approach to quantum evolution was first employed in polarization optics in the design of achromatic (broadband) polarization retarders where several wave plates are rotated at specific angles with respect to the fast polarization axis \cite{Westland}. Some thirty years later, the composite pulse approach was developed in NMR for design of composite pulse sequences.

The first composite pulse sequence for use in NMR was proposed by Levitt and Freeman \cite{Levitt79} to compensate for fluctuations in the radiofrequency driving strength.
This type of sequence which realizes a transition probability nearly constant over a wide range of pulse area variations around $\pi$ is called a broadband (BB) composite pulse.
The longer the composite sequences, the flatter and broader BB excitation profiles are obtained.
Alternatively, a narrow excitation profile around pulse area $\pi$ is produced by narrowband (NB) composite pulse sequences \cite{Tycko84,Tycko85,Shaka84,Wimperis89}.
Such NB pulses enable enhanced selectivity of excitation and make possible the selective addressing of an atomic qubit in a string of closely spaced qubits with a laser beam which is not tightly focused on the addressed qubit alone, but residual light may irradiate the adjacent qubits \cite{Ivanov2}.

Passband (PB) sequences exhibit both narrowband and passband behavior simultaneously and possess a nearly rectangular excitation profile as a function of the pulse area. They act as error correcting composite pulses for moderate deviations $\epsilon$ of the pulse area from $\pi$ and also as relatively robust identity operators for pulse areas near 0 and $2\pi$. This makes them an ideal tool for background suppression of unwanted signal in NMR experiments as was recently demonstrated in Ref. \cite{Odedra12} where improved intensity in the desired signal was achieved while still maintaining good suppression of unwanted background signal. Such pulses are of significant interest also for trapped ions or ultracold atoms in optical lattices for PB pulses ensure both robust manipulation and local qubit addressing.

In this paper, we present an analytical recipe for deriving arbitrarily accurate completely compensating resonant composite pulses against pulse amplitude noise. The pulse design which we present in this paper is based on concatenation of broadband $\B$ and narrowband $\N$ pulse sequences, composed of $N_b$ and $N_n$ pulses respectively, both of which are defined by analytical formulas for arbitrary odd number of pulses. In this manner, we can create arbitrarily large and arbitrarily accurate PB sequences of $\Nb\times \Nn$ pulses, with $\Nb$ and $\Nn$ both odd numbers. The phases of the constituent resonant pulses are given by a simple analytic formula, and so is the excitation profile, which allows one to explicitly assess the accuracy of these PB sequences.

\section{Broadband and narrowband composite pulse sequences}

The dynamics of a coherently driven two-state quantum system $|\psi_1\rangle \leftrightarrow |\psi_2\rangle$ is governed by the Schr\"{o}dinger equation
\be
\mathrm{i}\hbar \partial_t \mathbf{c}(t) = \mathrm{H}(t) \mathbf{c}(t),
\ee
for the probability amplitudes vector $\mathbf{c}(t) = [c_1(t),c_2(t)]^{\textnormal{T}}$.
The Hamiltonian of the system reads
\be
\mathrm{H}(t) = (\hbar/2) \Omega (t) \mathrm{e}^{-\mathrm{i}D(t)} |\psi_1\rangle\langle\psi_2| + \textnormal{H.c.},
\ee
where $\Omega(t)$ is the Rabi frequency and $D(t) = \int_{t_i}^{t} \Delta(t')dt'$ with $\Delta = \omega_0 - \omega$ being the detuning of the laser frequency $\omega$ and the Bohr transition frequency of the qubit $\omega_0$.
The unitary time-evolution propagator for the system $\mathrm{U}(t_f,t_i)$ takes the probability amplitude vector at the beginning of the interaction $\mathbf{c}(t_i)$ to the end of the interaction according to $\mathbf{c}(t_f) = \mathrm{U}(t_f,t_i) \mathbf{c}(t_i)$.
The propagator $\mathrm{U}(t_f,t_i)$ can be parameterized by two complex Cayley-Klein parameters $a$ and $b$ obeying $\vert a \vert^2 + \vert b \vert ^2 = 1$.
The population inversion probability for a system initially in state $|\psi_1\rangle$ $(c_1(t_i)=1,c_2(t_i)=0)$ is given by $|U_{12}|^2$.
For resonant driving ($\Delta =0$), which we assume throughout, $a$ and $b$ depend only on the pulse area $A(t) = \int_{t_i}^{t_f} \Omega(t')dt'$ through $a = \cos(A/2)$ and $b = -\mathrm{i}\sin(A/2)$, which produce Rabi oscillations of the population between the two states $|\psi_1\rangle$ and $|\psi_2\rangle$. However, this excitation profile is sensitive to pulse imperfections such as pulse duration and pulse intensity.
For example, a variation $\epsilon$ from the desired value for the pulse area results in an error in the population inversion of the order $\mathcal{O} (\epsilon^2)$.

A constant phase shift in the driving field, $\Omega (t) \rightarrow \Omega(t) \mathrm{e}^{\mathrm{i}\phi}$, produces the phased propagator
\begin{equation}
\mathbf{U}(\phi) = \left[ \begin{array}{cc}
a & b \mathrm{e}^{\mathrm{i}\phi} \\
-b^* \mathrm{e}^{-\mathrm{i}\phi} & a^* \end{array} \right].
\end{equation}
Then, the propagator of the two-state system subjected to a sequence of $N$ identical pulses with equal pulse areas $A$ and different phases $\phi_k$ of the driving fields is given by
\begin{equation}
\mathbf{U}^{(N)}(A) = \mathbf{U}(\phi_N)\cdots \mathbf{U}(\phi_2)\mathbf{U}(\phi_1).
\label{UN}
\end{equation}
(The operators act from right to left.) The assumption of equal pulse areas and different phases is natural for pulsed lasers because they produce a sequence of possibly imperfect, but identical pulses.
Each pulse can then be given a different phase shift by using an acousto-optical modulator, electro-optical modulator or a pulse shaper \cite{Weiner,Wollenhaupt05,Wollenhaupt06,Wollenhaupt07}.
From Eq. (\ref{UN}) we find that the population inversion probability for a system initially in state $|\psi_1\rangle$ is then given by $|U^{(N)}_{12}|^2$.
For a target Rabi angle $\pi$ and resonant pulses it is most convenient to use an odd number of pulses $N$, as in most of the existing literature; we make this assumption here too.

The composite phases $\phi_k$ ($k=1,2,\ldots,N$) are free control parameters, which are determined by the desired inversion profile. BB pulses are designed to produce transition probability $p \approx 1$ for pulse areas around $A=\pi$ (a flat excitation top), while NB pulses minimize $p \approx 0$ for pulse areas around $A=0$ (a flat excitation bottom). The PB sequences combine these effects and they produce, generally, an arbitrary $\Nb$-fold flat top at $A=\pi$ suppressing errors in the field Rabi amplitude in quantum optics (or the rf amplitude in NMR and the phase retardation in polarization optics) to order $\mathcal{O}(\epsilon^{\Nb})$, and an $\Nn$-fold flat bottom at $A=0$ suppressing errors to order $\mathcal{O}(\epsilon^{\Nn})$.

\emph{BB pulse sequences.} $\B$ pulse sequences of an odd number of pulses $\Nb$ for a target pulse area $\pi$ can be constructed by solving the set of algebraic equations \cite{Torosov},
\be \label{bb-cond}
[ U_{11}^{(\Nb)} ]_{A=\pi} = 0, \quad
[ \partial_A^k U_{11}^{(\Nb)} ]_{A =\pi} = 0,
\ee
for $k = 1,2,\ldots,\Nb$. This ensures an $\Nb$-fold flat-top profile around $A=\pi$. The BB composite phases are given by the analytic formula \cite{Torosov, TorosovPRL}
\begin{equation}\label{phases-BB}
\phaseB_k = \left(\Nb + 1 - 2 \left\lfloor \frac{k+1}{2} \right\rfloor \right) \left\lfloor \frac{k}{2} \right\rfloor \frac{\pi}{\Nb} ,
\end{equation}
where $\lfloor x \rfloor$ is the integer part of $x$ (the floor function).
For example, the first few BB composite sequences are $\B_3 = \{0, \tfrac23, 0\}\pi$, $\B_5 = \{0, \tfrac45, \tfrac25, \tfrac45, 0\}\pi$, and $\B_7 = \{0, \tfrac67, \tfrac47, \tfrac87, \tfrac47, \tfrac67, 0\}\pi$.

The inversion profiles around $A=\pi$ produced by these composite $\B$ pulses are given by \cite{Torosov,TorosovPRL}
\be\label{U11-bb}
U_{11}^{(\Nb)} = \cos^{\Nb}(A/2),
\ee
and hence the inversion probability is
\be\label{p-BB}
P^{(\Nb)} = 1-\cos^{2\Nb}(A/2) = 1- (1-p)^{\Nb},
\ee
with $p=\sin^{2}(A/2)$ being the single-pulse inversion probability.

\emph{NB pulse sequences.} $\N$ pulse sequences of $\Nn$ pulses can be constructed by solving the set of equations
\be \label{nb-cond}
[ U_{11}^{(\Nn)}]_{A=\pi} = 0, \quad [ \partial_A^j U_{11}^{(\Nn)} ]_{A =0} = 0,
\ee
for $j=1,2,\ldots,\Nn$. This ensures an $\Nn$-fold flat-bottom profile around $A=0$. The solution for the $\Nn$ NB phases is given by the analytic formula \cite{Vitanov}
\begin{equation}
\phaseN_j = (-)^j \left\lfloor \frac{j}{2} \right\rfloor \frac{2\pi}{\Nn}.
\label{phases-NB}
\end{equation}
Explicitly, the first few NB composite sequences are $\N_3=\left\{0,\tfrac{2}{3},-\tfrac{2}{3}\right\}\pi$, 
$\N_5=  \left\{0,\tfrac{2}{5},-\tfrac{2}{5},\tfrac{4}{5},-\tfrac{4}{5}\right\}\pi$, and 
$\N_7=\left\{0,\tfrac{2}{7},-\tfrac{2}{7},\tfrac{4}{7},-\tfrac{4}{7},\tfrac{6}{7},-\tfrac{6}{7}\right\}\pi.$
These composite $\N$ pulses produce inversion profiles around $A=\pi$ given by
\be
U_{12}^{(\Nn)} = e^{-i\pi/(2\Nn)} \sin^{\Nn}(A/2),
\ee
and therefore the excitation probability is
\be\label{p-NB}
P^{(\Nn)} = \sin^{2\Nn}(A/2) = p^{\Nn}.
\ee

The significance of the analytic formulas for BB and NB composite pulses, Eqs.~\eqref{phases-BB} and \eqref{phases-NB}, is in the fact that they define \emph{arbitrarily accurate} BB and NB inversion profiles because they are applicable for any (odd) number of pulses. This feature makes them substantially different from earlier BB and NB pulses, which achieved scalability by concatenating themselves, thereby producing composite sequences typically of $3^N$ and $5^N$ pulses. The analytic BB and NB sequences of Eqs.~\eqref{phases-BB} and \eqref{phases-NB} allow a much greater flexibility because they are applicable to any odd number of pulses rather than integer powers of 3 and 5 only.

\section{Design of composite passband pulses}

\subsection{Passband composite sequences}

In analogy with BB and NB pulses, the phases for a PB sequence of $N$ pulses of overall area $A = \pi$ are derived from the conditions
\bse\label{PB-eqs}
\begin{align} \label{condi}
\left[ U_{11}^{(N)} \right]_{A=\pi} &= 0, \\
\left[ \partial_A^k U_{11}^{(N)} \right]_{A =\pi} &= 0 \quad (k = 1,2,\ldots,\Nb), \label{api} \\
\left[ \partial_A^j U_{11}^{(N)} \right]_{A = 0} &= 0 \quad (j=1,2,\ldots,\Nn). \label{a0}
\end{align}
\ese
Note that equations \eqref{api} for even $k$ and Eqs.~\eqref{a0} for odd $j$ are fulfilled identically. Equations \eqref{PB-eqs} can be solved numerically by aiming at the largest possible values of the BB order $\Nb$ and NB order $\Nn$. The value of $\Nb$ controls the flatness of the top of the profile and the value of $\Nn$ controls the flatness of its bottom; these two numbers provide some flexibility of the PB profile. Numerical solutions, naturally, have some upper limits of the maximum values of $\Nb$ and $\Nn$, depending on the computing power.

\subsection{Nested $\N(\B)$ and $\B(\N)$ passband composite sequences}


Here we construct arbitrarily flexible and arbitrarily accurate PB composite pulses by concatenating the $\B$ and $\N$ sequences with the analytic phases of Eqs.~\eqref{phases-BB} and \eqref{phases-NB}.
We say that two prime composite sequences with phases $\N=\{\phaseN_j\}_{j=1}^{\Nn}$ and $\B=\{\phaseB_k\}_{k=1}^{\Nb}$ are concatenated to form a new sequence $\N(\B)$ if each constituent pulse of the prime $\N$ sequence is replaced by the prime $\B$ sequence (we say that $\B$ is nested into $\N$).
This is usually written as
\be
\N(\B) = \{\B_{\phaseN_1},\B_{\phaseN_2},\ldots,\B_{\phaseN_{\Nn}}\},
\label{nbpulse}
\ee
wherein the phase in each subscript is added to all phases in the $\B$ sequence. Hence the phases of the concatenated PB pulse $\N(\B)$ satisfy,
\be
\N(\B) :\quad \phi_{\Nb(j-1)+k} = \phaseN_j + \phaseB_k,
\label{nb}
\ee
with $j=1,2,\ldots,\Nn$ and $k=1,2,\ldots,\Nb$.
Examples of the several PB pulses $\N(\B)$ are presented in Table \ref{table1}.

\begin{table}[t]
\caption{Phases of several PB composite sequences $\N(\B)$ of different lengths and target area $\pi$.}
\begin{tabular}{l l}
\hline\noalign{\smallskip}\smallskip
$\N(\B)$ & $\{\phi_1, \phi_2, \ldots , \phi_N\}$ \\ \hline\noalign{\smallskip}\smallskip
$N_3(B_3)$ & $\{ 0, \frac23, 0, \frac23, \frac43, \frac23, \frac43, 0, \frac43 \}\pi$ \\ \noalign{\smallskip}
$N_3(B_5)$  & $\{ 0, \frac45, \frac25, \frac45, 0, \frac23, \frac{22}{15}, \frac{16}{15}, \frac{22}{15}, \frac23, \frac{4}{3}, \frac{2}{15}, \frac{26}{15}, \frac{2}{15}, \frac{4}{3} \}\pi$ \\ \noalign{\smallskip}
$N_3(B_7)$  & $\{ 0, \frac67, \frac47, \frac87, \frac47, \frac67, 0, \frac23, \frac{32}{21}, \frac{26}{21}, \frac{38}{21}, \frac{26}{21}, \frac{32}{21}, \frac23,$\\
& $ \frac43, \frac4{21}, \frac{40}{21}, \frac{10}{21}, \frac{40}{21}, \frac4{21}, \frac43\}\pi$ \\ \noalign{\smallskip}
$N_5(B_3)$  & $\{ 0, \frac23, 0, \frac25, \frac{16}{15}, \frac25, \frac{8}{5}, \frac{4}{15}, \frac{8}{5}, \frac45, \frac{22}{15}, \frac45, \frac{6}{5}, \frac{28}{15}, \frac{6}{5} \}\pi $\\ \noalign{\smallskip}
$N_5(B_5)$  & $ \{ 0, \frac{4}{5}, \frac{2}{5}, \frac{4}{5}, 0, \frac{2}{5}, \frac{6}{5}, \frac{4}{5}, \frac{6}{5}, \frac{2}{5}, \frac{8}{5},\frac{2}{5},0,\frac{2}{5},\frac{8}{5},\frac{4}{5},\frac{8}{5},\frac{6}{5},\frac{8}{5},\frac{4}{5},$ \\
 & $\frac{6}{5},0,\frac{8}{5},0,\frac{6}{5} \}\pi $ \\ \noalign{\smallskip}
$N_7(B_3)$  & $\{ 0,\frac{2}{3},0,\frac{2}{7},\frac{20}{21},\frac{2}{7},\frac{12}{7},\frac{8}{21},\frac{12}{7},\frac{4}{7},\frac{26}{21},\frac{4}{7},\frac{10}{7},\frac{2}{21},$ \\
& $\frac{10}{7},\frac{6}{7},\frac{32}{21},\frac{6}{7},\frac{8}{7},\frac{38}{21},\frac{8}{7}\}\pi$ \\ \noalign{\smallskip}
\hline 
\end{tabular}
\label{table1}
\end{table}

\begin{figure}[t]
\includegraphics[width=1\columnwidth]{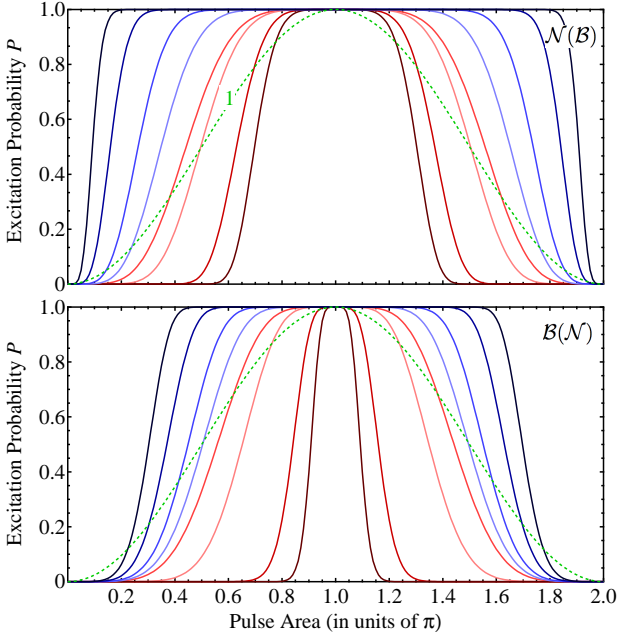}
\caption{(Color online) Excitation probabilities $P$ for PB composite pulses versus the area of the ingredient pulses. Green dotted curve shows the profile of a single $\pi$-pulse for easy reference. (top) Inversion profiles of $\N(\B)$ pulses with phases given by Eq. \eqref{nb}. The pulses from outside to inside are as follows, $N_3(B_{75})$, $N_3(B_{25})$, $N_3(B_{9})$, $N_3(B_{5})$, $N_3(B_{3})$, $N_5(B_{3})$, $N_{25}(B_{3})$, $N_{75}(B_{3})$.
(bottom) $\B(\N)$ pulses obtained by using the same ingredient NB and BB sequences as top frame but different concatenating approach \eqref{bnpulse} with phases given by Eq. \eqref{bn}. The pulses from outside to inside are as follows, $B_{75}(N_3)$, $B_{25}(N_3)$, $B_{9}(N_3)$, $B_{5}(N_3)$, $B_{3}(N_3)$, $B_{3}(N_{5})$, $B_{3}(N_{25})$, $B_{3}(N_{75})$.
}
\label{fig1}
\end{figure}

Alternatively, we can concatenate an $\N$ pulse into a $\B$ pulse. However, as noted before \cite{Levitt83,Shaka84,Wimperis89}, asymmetric pulse sequences, such as the NB sequences \eqref{phases-NB}, should be nested into another composite sequence by alternating the sequence $\N=\{\phaseN_1,\phaseN_2,\ldots,\phaseN_{\Nn}\}$ and its reverse $\N^{-1}=\{\phaseN_{\Nn},\ldots,\phaseN_2,\phaseN_1\}$.
Therefore, for the passband pulse sequence we obtain
\be
\B(\N)=\{\N_{\phaseB_1},\N_{\phaseB_2}^{-1},\N_{\phaseB_3},\N_{\phaseB_4}^{-1},\ldots,\N_{\phaseB_{\Nb}}\}.
\label{bnpulse}
\ee
The phases of the $\B(\N)$ sequence read
\be
\B(\N) :\quad \phi_{\Nn(k-1)+j} = \left\{\begin{array}{ll}\phaseN_j + \phaseB_k, & (\text{for}\ k\ \text{odd}) \\ \phaseN_{\Nn+1-j} + \phaseB_k, & (\text{for}\ k\ \text{even}) \end{array}\right. .
\label{bn}
\ee
Several PB sequences obtained in this manner are shown explicitly in Table \ref{table2}.

Design of composite pulses by concatenation is a widely adopted approach in NMR  \cite{Wimperis94,Tycko85,Cho86} and QIP \cite{Jones13,Husain13}.
Concatenating a prime composite pulse into itself overcomes the problem of solving large systems of equations in order to find the phases which becomes computationally difficult very quickly
   and furthermore, produces only approximate numerical values for the phases.
The design of nested PB composite pulses $\N(\B)$ and $\B(\N)$, which we present here, produces exact simple rational values for the phases but moreover,
 it offers flexibility and control of the excitation profile and steepness, as shown below.

\begin{table}[t]
\caption{Phases of several PB composite sequences $\B(\N)$ of different lengths and target area $\pi$.}
\begin{tabular}{l l}
\hline \noalign{\smallskip}\smallskip
$\B(\N)$ & $\{\phi_1, \phi_2, \ldots , \phi_N\}$ \\ \hline\noalign{\smallskip}
$B_3(N_3)$ & $\{ 0,\frac23,\frac43,0,\frac43,\frac23,0,\frac23,\frac43 \}\pi$ \\ \noalign{\smallskip}
$B_3(N_5)$  & $\{0, \frac25, \frac{8}{5}, \frac45, \frac{6}{5}, \frac{28}{15}, \frac{22}{15}, \frac4{15}, \frac{16}{15}, \frac23, 0, \frac25, \frac{8}{5}, \frac45, \frac{6}{5} \}\pi$ \\ \noalign{\smallskip}
$B_3(N_7)$  & $\{ 0,\frac{2}{7},\frac{12}{7},\frac{4}{7},\frac{10}{7},\frac{6}{7},\frac{8}{7}, \frac23, \frac{20}{21},\frac{8}{21},\frac{26}{21},\frac{2}{21},\frac{32}{21},\frac{38}{21},$ \\
& $0,\frac{2}{7},\frac{12}{7},\frac{4}{7},\frac{10}{7},\frac{6}{7},\frac{8}{7}\}\pi$ \\ \noalign{\smallskip}
$B_5(N_3)$  & $\{0, \frac23, \frac{4}{3}, \frac2{15}, \frac{22}{15}, \frac45, \frac25, \frac{16}{15}, \frac{26}{15}, \frac2{15}, \frac{22}{15}, \frac45, 0, \frac23, \frac{4}{3}\}\pi$ \\ \noalign{\smallskip}
$B_5(N_5)$  & $\{0,\frac{2}{5},\frac{8}{5},\frac{4}{5},\frac{6}{5},0,\frac{8}{5},\frac{2}{5},\frac{6}{5},\frac{4}{5},\frac{2}{5},\frac{4}{5},0,\frac{6}{5},\frac{8}{5},0,\frac{8}{5},\frac{2}{5},\frac{6}{5},\frac{4}{5},$ \\
& $0,\frac{2}{5},\frac{8}{5},\frac{4}{5},\frac{6}{5} \}\pi $ \\ \noalign{\smallskip}
$B_7(N_3)$  & $\{ 0,\frac{2}{3}, \frac{4}{3},\frac{4}{21},\frac{32}{21},\frac{6}{7},\frac{4}{7},\frac{26}{21},\frac{40}{21},\frac{10}{21},\frac{38}{21},\frac{8}{7},\frac{4}{7},\frac{26}{21},$ \\
& $\frac{40}{21},\frac{4}{21},\frac{32}{21},\frac{6}{7},0,\frac{2}{3},\frac{4}{3} \}\pi $ \\ \noalign{\smallskip}
\hline
\end{tabular}
\label{table2}
\end{table}

\subsection{Inversion profiles}

We proceed to derive analytic expressions for the inversion profiles of the concatenated PB sequences by using the known profiles of the $\B$ and $\N$ sequences presented above, Eqs.~\eqref{p-BB} and \eqref{p-NB}.
For example, the inversion profile of the $\N(\B)$ sequence is obtained by replacing the single-pulse transition probability $p$ in the NB inversion profile \eqref{p-NB} by the BB transition probability \eqref{p-BB}.
Similarly, the inversion profile for the PB sequence $\B(\N)$ is obtained by replacing the single-pulse transition probability $p$ in the BB inversion profile \eqref{p-BB} by the NB transition probability \eqref{p-NB}.
Thereby, we find
\bse
\label{p}
\begin{align}
P_{\N(\B)} &= \left( 1-(1-p)^{\Nb}\right) ^{\Nn},\\
P_{\B(\N)} &= 1-\left( 1-p^{\Nn}\right) ^{\Nb}.
\end{align}
\ese
 with $p = \sin^{2}(A/2)$.

Figure \ref{fig1} shows the excitation profile versus the area of the ingredient pulses for several $\B(\N)$ and $\N(\B)$ PB composite sequences whose phases are determined by Eqs.~\eqref{bn} and \eqref{nb}, respectively. 
This figure demonstrates the flexibility and huge range of the inversion profiles that can be produced by the nested PB pulses derived here. In agreement with Eqs. \eqref{p} the $\B(\N)$ concatenated pulses deliver narrower excitation profiles than $\B(\N)$ pulses. 
By increasing the length of the PB sequences we can tailor the transition probability $P$ in any desired manner.

\begin{figure}[t]
\includegraphics[width=1\columnwidth]{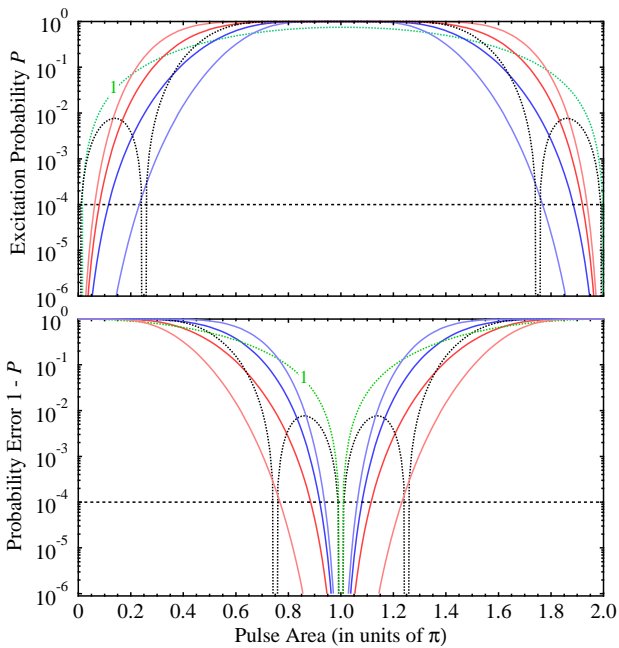}
\caption{(Color online) Fidelity of excitation profiles against the $10^{-4}$ benchmark error in quantum information for 9- and 15-pulse PB composite sequences of Eqs.~\eqref{nbpulse} and \eqref{bnpulse}, given explicitly in Tables \ref{table1} and \ref{table2}, versus the pulse area of the ingredient pulses. From outside to inside the PB pulses (depicted with solid curves) are: $N_3(B_5)$, $N_3(B_3)$, $B_3(N_3)$, $B_3(N_5)$. The dotted curve shows the profile of a single $\pi$-pulse for easy reference. The dashed curve is the inversion profile for the PB$_2$($\pi$) pulse of Wimperis \cite{Wimperis94}: $\{0,\frac12,\frac12,\frac{11}8,\frac{11}8,\frac{11}8,\frac{11}8,\frac12,\frac12\} \pi$. (top) Excitation probability $P$ (fidelity). (bottom) Excitation probability error $1-P$.}
\label{fig2}
\end{figure}

Figure \ref{fig2} shows the accuracy of the inversion profiles by using a logarithmic scale. As an accuracy benchmark we use the value $10^{-4}$, which is a popular benchmark error in quantum computation \cite{Nielsen}. A significant increase of the high-fidelity ranges compared to a single pulse is observed. This figure shows that, unlike the earlier 9-pulse PB$_2$($\pi$) pulse of Wimperis \cite{Wimperis94}, our 9-pulse PB sequences have both high-fidelity inversion around pulse area $\pi$ and high-fidelity suppression of excitation around pulse areas $0$ and $2\pi$.

\begin{figure}[t]
\includegraphics[width=1\columnwidth]{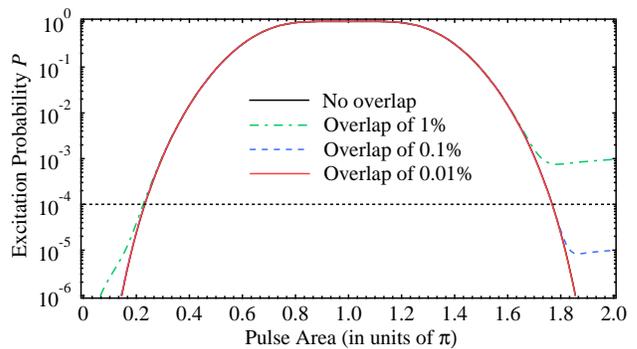}
\caption{(Color online) Excitation probability $P$ in log scale for the pulse sequence $B_3 (N_5)$ as a function of the pulse area for given temporal overlap between the individual pulses of 1\%, 0.1\% and 0.01\%.
For no overlap and overlap of 0.01\% the two curves lie on top of each other (solid lines).}
\label{fig3}
\end{figure}

Until now we assumed that the constituting pulses do not overlap in time, which is expressed in the decomposition of the quantum evolution in Eq.~{\eqref{UN}}.
However, real laser pulses can have "pre-pulses" or can decay slowly, which would lead to an overlap with the neighboring pulses in the sequence.
In order to evaluate the sensitivity of the excitation probability $P$ to pulse overlap we calculated $P$ for the pulse sequence $B_3(N_5)$ for temporal overlaps of 1\%, 0.1\% and 0.01\%. The results are shown in Fig. \ref{fig3} where it is easy to see that population inversion is robust against such errors.

\begin{figure}[t]
\includegraphics[width=1\columnwidth]{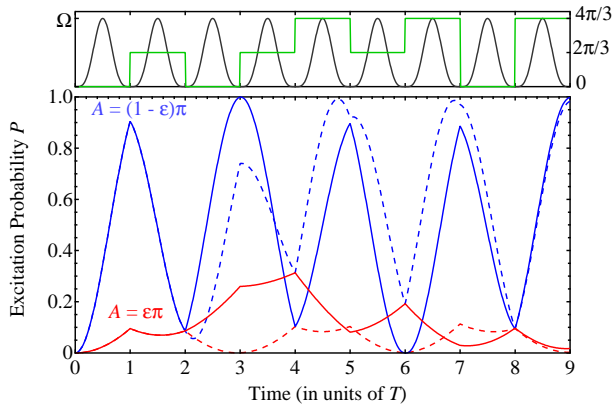}
\caption{(Color online) (top) A train of identical pulses with maximum Rabi frequency $\Omega$ and phases $0$, $2\pi/3$ or $4\pi/3$ represented by the digital curve corresponding to composite sequence $N_3 (B_3)$.
(bottom) Excitation probability $P$ as a function of time in units of $T$ (pulse duration of each constituent pulse) for the nine-pulse composite sequences $N_3 (B_3)$ (solid lines) and $B_3 (N_3)$ (dashed lines). Each constituent pulse is assumed to have an intrinsic amplitude error $\varepsilon = 0.2$. The blue curves represent $P$ for pulse area $A = (1-\varepsilon) \pi$ while the bottom curves depict $P$ for pulse area $A = \varepsilon \pi$. At the end of the pulse sequence, due to destructive interference of the errors, complete population inversion ($P = 1$) is achieved for $A=(1-\varepsilon)\pi$, while complete population return ($P = 0$) is achieved for $A=\varepsilon \pi$. }
\label{fig4}
\end{figure}

In order to elucidate the underlying quantum evolution during a composite pulse sequence we present an illustration of the population evolution in Fig. {\ref{fig4}}. We chose the 9-pulse sequences $N_3 (B_3)$ and $B_3 (N_3)$, in which each constituent pulse has an area $A = (1-\varepsilon) \pi$, with $\varepsilon=0.2$. A single pulse with pulse area $A = (1-\varepsilon) \pi$ produces an excitation probability 0.9 while a pulse with area  $A = \varepsilon \pi$ produces  excitation probability 0.1. When we use composite pulse technique, at the end of the pulse sequence destructive interference of the pulse errors drives the system to a complete population inversion.
The opposite behavior is observed for individual pulse area $A = \varepsilon \pi$, with $\varepsilon=0.2$: each pulse produces some small excitation but in the end of the composite pulse the destructive interference suppresses excitation completely.


\subsection{Width of the profiles}

Equations \eqref{p} allow us to readily derive the basic properties of the inversion profiles. The half-width-at-half-maximum (HWHM) (i.e. at $P_{\N(\B)}=\frac12$ and $P_{\B(\N)}=\frac12$) of the inversion profiles read $A_{\frac12} = \pi - 2\arcsin\sqrt{p_{\frac12}}$ where
\bse\label{p_1/2}
\begin{align}
\N(\B):&\quad p_{\frac12} = 1-\big( 1-2^{-1/\Nn}\big) ^{1/\Nb},\\
\B(\N):&\quad p_{\frac12} = \big( 1-2^{-1/\Nb}\big) ^{1/\Nn}.
\end{align}
\ese
For $\Nn\gg 1$ and $\Nb\gg 1$ we find
\bse\label{width-approx}
\begin{align}
\N(\B):&\quad A_{\frac12} \sim \pi - 2 \sqrt{\frac{\ln(\Nn/\ln2)}{\Nb}},\\
\B(\N):&\quad A_{\frac12} \sim 2 \sqrt{\frac{\ln(\Nb/\ln2)}{\Nn}}.
\end{align}
\ese
This shows that the HWHM of the $\N(\B)$ pulses is determined primarily by $\Nb$, while the HWHM of the $\B(\N)$ pulses is determined primarily by $\Nn$.

\subsection{Rectangularity (steepness) of the profiles}

Another important aspect of the PB composite sequences is the degree of ``rectangularity'' of the inversion profile, i.e. the steepness of the rise and fall of the population inversion. This steepness is described by the derivative of $P$ with respect to the individual pulse area $A$ at the mid-point $p_{\frac12}$ where $P=\frac12$, as given by Eqs.~\eqref{p_1/2}. The inverse value of this derivative, $\delta A = 1/[\partial_A {P}]_{P=\frac12}$ is a measure of the interval of pulse areas $\delta A$ over which $P$ rises from about 0.1 to about 0.9 \cite{Vitanov99,Boradjiev10}.

A simple calculation gives
\bse\label{steepness}
\begin{align}
\N(\B):&\quad \delta A = \frac{2}{\Nb\Nn \big(2^{\frac1{\Nn}}-1\big) \sqrt{\big(1-2^{-\frac1{\Nn}}\big)^{-\frac1{\Nb}}-1}} ,\\
\B(\N):&\quad \delta A = \frac{2 \big(1-2^{-\frac1{\Nb}}\big)^{\frac1{2\Nn}}}{\Nb\Nn \big(2^{\frac1{\Nb}}-1\big) \sqrt{1-\big(1-2^{-\frac1{\Nb}}\big)^{\frac1{\Nn}}}} .
\end{align}
\ese
For $\Nn\gg 1$ and $\Nb\gg 1$ we find
\bse\label{steepness-approx}
\begin{align}
\N(\B):&\quad \delta A \sim \frac{2/\ln2}{\sqrt{\Nb \ln(\Nn/\ln2)}},\\
\B(\N):&\quad \delta A \sim \frac{2/\ln2}{\sqrt{\Nn \ln(\Nb/\ln2)}}.
\end{align}
\ese
As for the width of the inversion profile, the steepness of the inversion profile for $\N(\B)$ pulses is determined primarily by $\Nb$, while for the $\B(\N)$ pulses it is determined primarily by $\Nn$.

\begin{figure}[t]
\includegraphics[width=1\columnwidth]{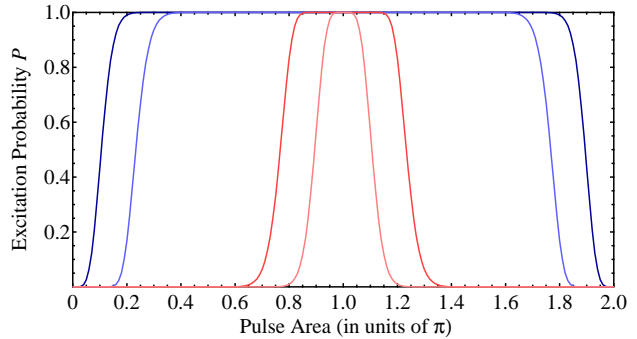}
\caption{(Color online) Excitation profiles with the same rectangularity $\delta A \approx 0.1\pi$ of $\N(\B)$ and $\B(\N)$ composite pulses with phases given, respectively, by Eqs.~\eqref{nb} and \eqref{bn}.
The pulses from outside to inside are as follows: $N_{3}(B_{57})$, $N_{21}(B_{25})$, $B_{21}(N_{25})$, $B_{3}(N_{57})$.}
\label{fig5}
\end{figure}

These simple analytic formulas allow us to estimate the needed resources for producing any desired PB inversion profile.
For example, Fig.~\ref{fig5} shows a set of inversion profiles for various PB sequences, which have the same, pre-selected rectangularity $\delta A\approx 0.1\pi$.
The needed number of pulses was calculated in each case by using the analytic formulas \eqref{steepness-approx}.

\subsection{Experimental feasibility}

Composite pulse sequences like the ones presented in this paper can be experimentally realized in various ways.
They are routinely produced in the radiofrequency and microwave domains by appropriate generators, which easily produce pulses of any shape and phase and any delay.
In the optical domain, such sequences are commonly known as pulse trains.
Sequences of pulses of microsecond duration can be produced from a cw laser beam by mechanical choppers and acousto-optical modulators.
A nanosecond pulse train can be produced by a nanosecond laser system with an appropriate repetition rate.
The needed phase shift can be imposed by electro-optical or acousto-optical modulators.
For microsecond and nanosecond pulses, care must be taken to make sure that there is no decoherence during the composite pulse sequence; hence, one should use typically (two-photon) transitions between ground or metastable atomic states.
Pico- and femtosecond pulses can be produced by femtosecond laser systems \cite{Weiner,Wollenhaupt05,Wollenhaupt06,Wollenhaupt07}.
A train of femtosecond pulses can be produced by a laser operating at, e.g., 100 MHz repetition rate;
 in this case, even a transition between a ground and a long-lived excited state can be used because the delay between the pulses in just a few nanoseconds.
A train of femtosecond pulses, with a temporal delay in the femtosecond range too, can be generated by pulse-shaping technologies \cite{Weiner,Wollenhaupt05,Wollenhaupt06,Wollenhaupt07};
 on such times scales, of course, no decoherence takes place.

\section{CONCLUSION}

We presented a general method for the design of arbitrarily large and arbitrarily accurate PB composite pulses whose phases are given by simple analytical formulas, Eqs.~\eqref{nb} and \eqref{bn}. 
Our approach is based on designing the SU(2) propagator of the system by concatenating broadband and narrowband pulse sequences. 
We construct two types of PB pulses: $\N(\B)$ in which a broadband pulse is nested into a narrowband pulse, and $\B(\N)$ in which a narrowband pulse is nested into a broadband pulse. 
The $\N(\B)$ inversion profiles are wider than the $\B(\N)$ profiles, which provides some leeway of choice. 
In this manner, arbitrary chosen inversion profiles can be generated that possess different arbitrary flat top and bottom properties. 
An important advantage of our method is that it can produce any PB composite pulse sequence comprising of odd-number constituent pulses.  
For comparison, existing proposals are based on numerical methods for finding the phases which greatly limits their scalability, or on concatenating a basic PB pulse with itself which leads to exponential growth of the number of pulses involved.
Furthermore, the PB composite pulses presented in this paper produce excitation profiles in which the robustness against variations in the parameters is accompanied with ultrahigh fidelity, well beyond the fault tolerance limit of quantum computation \cite{Nielsen}.
Unlike all earlier PB pulses, we derive simple exact analytic formulas for the inversion profiles, as well as for their width and steepness, which allow an easy estimation of the PB sequence needed for a desired population inversion profile.


\acknowledgments
E.K.~acknowledges financial support from SUTD start-up grant SRG-EPD-2012-029 and SUTD-MIT International Design Centre (IDC) grant IDG31300102.
N.V.V. acknowledges support from the EC Seventh Framework Programme under grant agreement No. 270843 (iQIT).


\end{document}